\documentstyle[12pt]{article}
\begin{document}
\title{TWO-PARAMETRIC QUANTUM-DEFORMED DUAL AMPLITUDE }
\vspace{5 in}
\author{\bf{L.Jenkovszky}\\
 Bogoliubov Institute for Theoretical Physics\\
 Metrologicheskaya St. 14b, 252143, Kiev-143, Ukraine.\\
 \bigskip
 E-mail:jenk@gluk.apc.org\\
 \bf{M.Kibler}\\
 Institut de Physique Nucl\'eaire de Lyon\\
 IN2P3-CNRS et Universit\'e Claude Bernard\\
 43 Boulevard du 11 Novembre 1918\\
 F69622 Villeurbanne Cedex, France\\
 \bigskip
 E-mail:kibler@frcpn11.bitnet\\
 \bf{A.Mishchenko}\\
 Bogoliubov Institute for Theoretical Physics\\
 Metrologicheskaya St. 14b, 252143, Kiev-143, Ukraine.\\
 E-mail:yusitenko@gluk.apc.org\\}
\maketitle
\begin{abstract}
 From a 2-parametric deformation of the harmonic oscillator
algebra we construct a 4-point dual amplitude with nonlinear trajectories.
The earlier versions of the q-deformed dual models are reproduced as
limiting cases of the present model.
\end{abstract}
\newpage

        Originally, dual models were thought to be promising
examples of hadronic scattering amplitudes in the framework
of the $S$-matrix theory. Subsequently, they gave rise to many formal
and mathematical developments while the original goal of constructing
realistic scattering amplitudes was abandoned. One of the
difficult problems that remained unsolved is that of the spectrum of particles
in the model and the related problem of the form of the Regge
trajectories used.

We remind that in the original Veneziano amplitude [1] the
trajectories are linear and the model predictes the existence of an
infinite sequence of infinitely narrow "resonances" whose spin
increases like their squared masses. Narrow resonance dual models
have the virtue of being factorizable [2], giving rise to a
field-theoretical interpretation of the scattering amplitude in terms
of relevant verticies and propagators, which in principle can be used
in the construction of a unitarization procedure (calculation of loop
diagrams) based on the Veneziano dual amplitude as the "Born term".
In spite of many efforts with partial successes, this programm has
never been completed.

A different class of dual models [3-5] is based on logarithmic
trajectories. The existence of ghosts -
states with negative norm - was one of the main reason
why the study of this second class of dual models was abandoned.

Factorizability made it possible for both classes
of the dual models to construct an operator formalism corresponding
to an infinite set of oscillators. It has been also realized
that the algebra of the dual models with logarithmic
trajectories can be viewed as the $q$-deformed algebra of harmonic
oscillators (Veneziano model).

Recent interest in quantum algebras (q-deformed algebras) brought
also a revival to the traditional dual models. Various
$q$-deformations of the harmonic oscillators and the resulting "dual
amplitudes" have been studied in a number of papers [4,6,7].

  In the present paper we
 construct a quantum-deformed dual amplitude by means of a
 2-parametric deformation of the dual amplitude.

 We remind that a similar problem in the case of a single-parameter
 deformation was considered in refs. [4,6,7]. The aim of
 all these attempts was to replace the infinite-dimensional
 algebra of the independent harmonic oscillators by a $q$-deformed
 quon algebra [8] which essentially means the introduction of some
 anharmonism in the interaction [9]. Correspondingly are the vertices
 and the propagators also modified.

 In refs. [4,7], the folowing deformation was used:
 $$a_\mu^{(n)}a_\nu^{+(n)}-qa_\nu^{+(n)}a_\mu^{(n)}=-\eta_{\mu\nu}, \quad
 n=\overline{1,\infty},\eqno(1)$$
 where $\eta_{\mu\nu}=diag(1,-1,...,-1), \mid{q}\mid<1$.

 In ref. [6] on the other hand, the $q$ deformation was defined as
$$a_\mu^{(n)}a_\nu^{+(n)}-qa_\nu^{+(n)}a_\mu^{(n)}=-\eta_{\mu\nu}q^{-N_n},$$
$$[N_n,a_\mu^{(n)}]= -a_\mu^{(n)}, \quad
[N_n,a_\mu^{+(n)}]= a_\mu^{+(n)}. \eqno(2)$$

 Note that in eqs. (1) and (2) $q$ is a real parameter and the
 interaction between various modes $n$ and $m$ is absent:
 $$[a_\mu^{(n)},a_\mu^{(m)}]=[a_\mu^{(n)},a_\mu^{+(m)}]=
 [a_\mu^{+(n)},a_\mu^{+(m)}]=0, \quad n\neq m.\eqno(3)$$

 The 2-parametric deformation to be inroduced in the present paper
 generalizes these two approaches and contains them as limiting
 cases.

 Let us remind that the 4-point dual Veneziano amplitude is defined
 by the following expression
 $$A_4(s,t)=B(-\alpha_0(s),-\alpha_0(t))=
 \int\limits_{0}^{1}dxx^{-\alpha_0(s)-1}(1-x)^{-\alpha_0(t)-1}$$
 $$= <0\mid V(p_2)\int\limits_{0}^{1}dx
 x^{H-\alpha_0(s)-1}(1-x)^{\alpha(0)-1}V(p_3)\mid0>,\eqno(4)$$
 where $\alpha_0$ is a (linear) Regge trajectory
$$\alpha_0(s)=\alpha^{\prime}s+\alpha(0), \quad
 p_i^2=m^2, \quad \alpha^{\prime}m^2+\alpha(0)=0,$$
$$2\alpha^{\prime}({p_2}{p_3})=\alpha^{\prime}t+2\alpha(0),$$
$$2\alpha^{\prime}({p_1}{p_2})=\alpha^{\prime}s+2\alpha(0),\eqno(5)$$
 depending on the
 invariant (Mandelstam) kinematical variables
$$s=({p_1}+{p_2})^2, \quad t=({p_2}+{p_3})^2, \quad u=({p_1}+{p_3})^2.$$
 The relevant vertex operator and the Hamiltonian are
$$V(p)=exp(-\sqrt{2\alpha^{\prime}}p^{\mu}
\sum_{n=1}^{\infty}\frac{a_{\mu}^{+(n)}}{\sqrt{n}})
exp(\sqrt{2\alpha^{\prime}}p^{\mu}\sum_{n=1}^{\infty}
\frac{a_{\mu}^{(n)}}{\sqrt{n}}),$$
$$H=-\sum_{n=0}^{\infty}na_{\mu}^{+(n)}a^{\mu(n)}\eqno(6)$$

 and the creation and annihilation operators
 $a_\mu^n, \quad a_\mu^{+(n)}$ in (6)
satisfy the  commutation relations (1) for $q$=1.

 For $\alpha (0)=1$, the amplitude (4) exhibits remarkable
 factorization properties, as one can easily verify by integrating
 (4) over $x$:
 $$A_4(s,t)=<0\mid V(p_2)[H-\alpha_0(s)]^{-1}V(p_3)\mid0>.\eqno(7)$$

  From (7) the pole structure of the amplitude can be easily found;
 the poles are located at
 $$\alpha_0(s_n)={\alpha^{\prime}}s_n+\alpha(0)=E_n, \quad
 s_n=\frac{E_n-\alpha(0)}{\alpha^{\prime}},\eqno(8)$$
 where $E_n$ is the eigenvalue of the operator $H$.

Crossing symmetry and duality imply [10,11]
 $$A_4(s,t)=-\frac{(\alpha_0(s)+\alpha_0(t))}{\alpha_0(s)\alpha_0(t)}
 \prod_{n=1}^{\infty}
 \frac{n[n-\alpha_0(s)-\alpha_0(t)]}{[n-\alpha_0(s)][n-\alpha_0(t)]}$$
 $$=-\sum_{n=0}^{\infty}\frac{(\alpha_0(s)+1)...(\alpha_0(s)+n)}{n!}
 \frac{1}{\alpha_0(t)-n}$$
 $$=A_4(t,s)=-\sum_{n=0}^{\infty}
 \frac{(\alpha_0(t)+1)...(\alpha_0(t)+n)}{n!}\frac{1}{\alpha_0(s)-n}.\eqno(9)$$

 In deforming the Veneziano amplitude $A_4(s,t)$, we use the
 following quon algebra [12]:
$$a_\mu^{(n)}a_\nu^{+(n)}-q_1a_\nu^{+(n)}a_\mu^{(n)}=
-\eta_{\mu\nu}q_2^{-N_n}, \quad \mu,\nu=\overline{1,d}, \quad
n=\overline{1,\infty}$$
$$[N_n,a_\mu^{(n)}]= -a_\mu^{(n)},\quad [N_n,a_\mu^{+(n)}]= a_\mu^{+(n)},$$
$$[a_\mu^{(n)},a_\nu^{(m)}]=[a_\mu^{(n)},a_\nu^{+(m)}]=
[a_\mu^{+(n)},a_\nu^{+(m)}]=0, \quad m\neq n,\eqno(10)$$
 where $q_1$ and $q_2$ are real numbers whose moduli do not belong to
 $[0,1[$ simultaneousely.

Introduce now the following notations, usefull in what follows
$${a^{(n)}}(p)={a_\mu^{(n)}}p^\mu, \quad
{a^{+(n)}}(p)={a_\mu^{+(n)}}p^\mu,$$
$$H_n=a_{\mu}^{+(n)}a^{\mu(n)},\eqno(11)$$
and
$$(x)_q=\frac{q_1^x - q_2^x}{q_1 - q_2},\eqno(12)$$
where $p^\mu$ is a c-number vector and $x$ is a number or
operator.

To derive the deformed amplitude, first we need the construction of
the representation of the quon algebra (10). The precedure is
standard and is based on the introduction of the vacuum (cyclic
vector $\mid0>)$:
$$a_{\mu}^{(n)}\mid0>=0, \quad N\mid0>=0,\eqno(13)$$
as well as the state vectors - by applying the creation operators
$a_{\mu}^{+(n)}$ on the vacuum.

We are interested only in Lorentz-invariant combinations (11), for
which the relations
$${a^{(n)}}(p_1){a^{+(n)}}(p_2)-q_1{a^{+(n)}}(p_2){a^{(n)}}(p_1)=
-(p_1p_2)q_2^{N_n},$$
$$H_n{a^{+(n)}}(p)\mid0>=-{a^{+(n)}}(p)\mid0>.\eqno(14)$$
hold. By induction, the following relations can be easily derived
 from (10) and (14):
$$H_n[{a^{+(n)}}(p)]^k\mid0>=-(k)_q[{a^{+(n)}}(p)]^k\mid0>$$
$$<0\mid [{a^{(n)}}(p_1)]^k[{a^{+(n)}}(p_2)]^l\mid0>=
\delta_{kl}(-1)^k[(k)_q!](p_1p_2)^k,\eqno(15)$$
where
$$(0)_q!=1, \quad (k)_q!=\prod_{j=1}^{k}(j)_q!, \quad k\geq1.\eqno(16)$$
The relations (15) will be basic when calculating the deformed
amplitude.

Our objective is the construction of a deformed dual and
factorizable scattering amplitude. Any factorizable reepresentation
for the scattering amplitude will require the calculation of the
vacuum expectation values of the form
\newpage
$$<0\mid V(a^{(n)}(p_1))D(H_n)V(-a^{+(n)}(p_2))\mid0>$$
$$=\sum_{n=0}^{\infty}V_1^{(n)}V_2^{(n)}[(n)_q!](p_1p_2)^nD(-(n)_q),\eqno(17)$$
where
$$V_i(x)=\sum_{n=0}^{\infty}V_i^{(n)}x^n.\eqno(18)$$
Note that eq. (17) is invariant under the replacement
$H_n\longrightarrow -(N_n)_q$.

Now, following [6,7] we define a new vertex operator replacing the
usual exponential by a deformed one
$$e_q(x)=\sum_{n=0}^{\infty}\frac{x^n}{[(n)_q!]},\eqno(19)$$
satisfying the equation
$$\frac{e_q(q_1x)-e_q(q_2x)}{x(q_1-q_2)}=e_q(x).\eqno(20)$$
It is used in the constructin of the coherent states for the
deformed oscillator algebra [13]
$$aa^+-q_1a^+a=q_2^{-N}.\eqno(21)$$
Really, it can be easily shown with the help of (20) and (21) that
$$a\mid z>=z\mid z>, \quad \mid z>=e_q(za^+)\mid0>.\eqno(22)$$
The new vertex is now
$$V(p)=\prod_{n=1}^{\infty}e_q(-\frac{\sqrt{2\alpha^{\prime}}a^{+(n)}(p)}
{\sqrt{n}})
e_q(\frac{\sqrt{2\alpha^{\prime}}a^{(n)}(p)}{\sqrt{n}}),\eqno(23)$$
It reduces to a non-deformed one as $q_1\rightarrow 1$ and
$q_2\rightarrow 1$.

Now by analogy with (6), we define
$$H=-\sum_{n=0}^{\infty}na_{\mu}^{+(n)}a^{\mu(n)}=-\sum_{n=0}^{\infty}nH_n,
\eqno(24)$$
and by using (17) we find
$$<0\mid V(p_2)x^HV(p_3)\mid0>=F(x,p_2p_3)$$
$$=\prod_{n=1}^{\infty}(\sum_{m=0}^{\infty}\frac{1}{(m)_q!}
(\frac{2\alpha^{\prime}}{n}(p_2p_3))^mx^{n(m)_q}).\eqno(25)$$
It can be easily verified that the insertion of (25) into (4)
violates crossing symmetry. To restore symmetry, we can either define
the propagator as [7]
$$D(p_1p_2)=\int\limits_{0}^{1}dxx^HF(1-x,p_1p_2),\eqno(26)$$
and
$$A_4=<0\mid V(p_2)D(p_1p_2)V(p_3)\mid0>,\eqno(27)$$
or, following ref. [6] define a set of the operators
$b_\mu^{(n)}, b_\mu^{+(n)}$ commuting
with $a_\mu^{(n)}, a_\mu^{+(n)}$ and similar
to $a_\mu^{(n)}, a_\mu^{+(n)}$
in the sense of the mutual commutation
relations (i.e. satisfying (10) under the replacement
$a_\mu^{(n)}\longrightarrow b_\mu^{(n)}$ and
$a_\mu^{+(n)}\longrightarrow b_\mu^{+(n)}$).

The next step towards the restoration of duality and crossing symmetry
is to use the representation [6]
$$A_4= \int\limits_{0}^{1}dx<0\mid V_a(p_2)V_b(p_1)x^{H_a}(1-x)^{H_b}
V_b(p_2)V_a(p_3)\mid0>,\eqno(28)$$
where the indeces $a$ and $b$ mean that in the corresponding
function the arguments are operators
$a_\mu^{(n)},a_\mu^{+(n)}$ or $b_\mu^{(n)},b_\mu^{+(n)}$.
The two approaches result in the same result for the deformed
amplitude
$$A_4(p_2p_3,p_1p_2)=\int\limits_{0}^{1}dxF(x,p_2p_3)F(1-x,p_1p_2),\eqno(29)$$

It can be seen from (25) that eq. (29) reduces to a non-deformed
amplitude (4) for $\alpha(0)=1$ in the limit $q_1\rightarrow 1$ and
$q_2\rightarrow 1$, to the one given in [6] when $q_1=q, \quad q_2=q^{-1}$ and
that of ref. [7] when $q_1=q, \quad q_2=1.$

Let us now find the position of the poles and the Regge trajectory for the
amplitude (29). To
this end one has to study the behaviour of the integrand in (29) and
consequently - the convergence of the infinite product (25)
defining the function $F(x,p_2p_3).$ Rewrite (25) in the form
$$F(x,p_2p_3)=\prod_{n=1}^{\infty}(1+c_n(x)),$$
$$c_n(x)=\sum_{m=1}^{\infty}\frac{1}{(m)_q!}
(\frac{2\alpha^{\prime}}{n}(p_2p_3))^mx^{n(m)_q},\eqno(30)$$
suitable for the utilisation of the familiar convergence criterion of
infinite products [11], according to which a product of the type (30)
converges absolutely if the series
$$\sum_{n=1}^{\infty}c_n(x)\eqno(31)$$
converges absolutely.

To study the convergence of the series (31) with the coefficients
$c_n(x),$ defined by (30), we first estimate the ratio
$$\frac{\mid{c_{n+1}(x)}\mid}{\mid{c_n(x)}\mid}=\frac{\sum_{m=1}^{\infty}
\frac{1}{(m)_q!}\mid{\frac{2\alpha^{\prime}}{n+1}(p_2p_3)}\mid
^mx^{(n+1)(m)_q}}
{\sum_{m=1}^{\infty}\frac{1}{(m)_q!}
\mid{\frac{2\alpha^{\prime}}{n}(p_2p_3)}\mid^m x^{n(m)_q}}$$
$$\leq \mid{x}\mid\frac{\sum_{m=1}^{\infty}\frac{1}{(m)_q!}\mid
{\frac{2\alpha^{\prime}}{n+1}(p_2p_3)}\mid^m x^{n(m)_q}}{\sum_{m=1}^{\infty}
\frac{1}{(m)_q!}\mid{\frac{2\alpha^{\prime}}{n}(p_2p_3)}
\mid^m x^{n(m)_q}} \leq \mid{x}\mid. \eqno(32)$$
It follows from (32) that the series (31), and consequently the
infinite product (30) converge absolutely for $x\in [0.1[$. Hence, the
integrand in (29) may have singularities only at the endpoints
of the integration $x=0$ and $x=1$.

To determine the behaviour of the function $F(x,p_2p_3)$ at
$x\rightarrow(1-0),$ we write for the first term in the
expansion in powers of $(p_2p_3)$:
$$lnF(x,p_2p_3)=-2\alpha^{\prime}(p_2p_3)ln(1-x) +
\sum_{k=2}^{\infty}[2\alpha^{\prime}(p_2p_3)]^kF_k(x).\eqno(33)$$
 From (33) it follows that
$$F(x,p_2p_3)\sim(1-x)^{-2\alpha^{\prime}(p_2p_3)}, \quad x \rightarrow (1-0).
\eqno(34)$$
Now, by using (25), (29) and (34), we find that the poles of the
amplitude (29) in $p_2p_3$ will be located at
$$2\alpha^{\prime}(p_2p_3)=n_0+1+\sum_{i=1}^{\infty}i(n_i)_q,\quad n_0,n_i
\in Z^+, \quad Z^+=0\cup N, \eqno(35)$$
where $N$ is the set of natural numbers. Here the term $n_0$ comes from the
expansion of
$$F(x,p_2p_3)(1-x)^{2\alpha^{\prime}(p_2p_3)},$$
in powers of $(1-x)$, the term 1 comes from the integration over $x$ in (29)
and the last term in (35) comes from $F(1-x,p_1p_2)$.

The residue at the pole (35) is a polynomial of power
$$\sum_{i=1}^{\infty}n_i, \eqno(36)$$
in $(p_1p_2)$ (with $n_i$ from (35)). Then we find from (35) and (36) the
leading Regge trajectory in an implicit form
$$(J)_q = \frac{q_1^J - q_2^J}{q_1 - q_2}=m^2-1, \eqno(37)$$
or (see also [14])
$$(\alpha(t))_q = \frac{q_1^{\alpha(t)} - q_2^{\alpha(t)}}{q_1 - q_2}=
\alpha^{\prime}t+2\alpha(0)-1. \eqno(38)$$

In the case $q_1=q, \quad q_2=q^{-1}$ we obtain from (37), (38) the following
spin-squared mass relation
$$J(m^2)=\frac{1}{lnq}ln\frac{[(q-q^{-1})(m^2-1)+
\sqrt{(q-q^{-1})^2(m^2-1)^2+4}]}{2},\eqno(39)$$
or Regge trajectory
$$\alpha(t)=\frac{1}{lnq}ln\frac{[(q-q^{-1})(\alpha^{\prime}t+2\alpha(0)-1)+
\sqrt{(q-q^{-1})^2(\alpha^{\prime}t+2\alpha(0)-1)^2+4}]}{2},\eqno(40)$$
while for $q_1=q, \quad q_2=1$ we get a logarithmic trajectory
$$J(m^2)=\frac{ln[(q-1)(m^2-1)+1]}{lnq}, \eqno(41)$$
or
$$J(m^2)=\frac{ln[(q-1)(\alpha^{\prime}t+2\alpha(0)-1)+1]}{lnq}, \eqno(42)$$
in agreement with [4].

The algebra (10) is not very convenient since
nothing can be said about the commutators of the type
$[a_\mu^{(n)},a_\nu^{(n)}]$ and
$[a_\mu^{+(n)},a_\nu^{+(n)}]$. For this reason, it is of interest
to find a deformed algebra
of the oscillators $d_\mu^{(n)},d_\mu^{+(n)}$ resulting in
the same amplitude for the
substitution
$a_\mu^{(n)} \rightarrow d_\mu^{(n)}, a_\mu^{+(n)} \rightarrow d_\mu^{+(n)}$
but with known comutators between the elements of
the algebra. It can be easily varified that the relevant algebra is
$$d_\mu^{(n)}d_\nu^{+(n)}-[\frac{(N_n+1)_q}{(N_n)_q}\frac{N_n}{N_n+1}]
d_\nu^{+(n)}d_\mu^{(n)}=
-\eta_{\mu\nu}\frac{(N_n+1)_q}{N_n+1},$$
$$[d_\mu^{(n)},d_\nu^{(n)}]=[d_\mu^{+(n)},d_\nu^{+(n)}]=0,$$
$$[N_n,d_\mu^{(n)}]=-d_\mu^{(n)},[N_n,d_\mu^{+(n)}]=
d_\mu^{+(n)}, \quad n=\overline{1,\infty}, \eqno(43)$$
(the operators corresponing to different modes $n\neq m$ commute)
since eq. (17) is invariant under the replacement
$a_\mu^{(n)} \rightarrow d_\mu^{(n)}, a_\mu^{+(n)} \rightarrow d_\mu^{+(n)}.$

Algebra (43) admits for a representation by means of the convential
(non-deformed) oscillators
$$d_\mu^{(n)}=\sqrt{\frac{(N_n+1)_q}{N_n+1}}{\tilde d}_\mu^{(n)},$$
$$d_\mu^{+(n)}={\tilde d}_\mu^{+(n)}\sqrt{\frac{(N_n+1)_q}{N_n+1}},\eqno(44)$$
where
$$[{\tilde d}_\mu^{(n)},{\tilde d}_\nu^{+(n)}]=-\eta_{\mu\nu},$$
$$[{\tilde d}_\mu^{(n)},{\tilde d}_\nu^{(n)}]=
[{\tilde d}_\mu^{+(n)},{\tilde d}_\nu^{+(n)}]=0,$$
$$ N_n={\tilde d}_\mu^{+(n)}{\tilde d}^{\mu n},\eqno(45)$$
and the opetors corresponding to various modes $n\neq m$ commuting among
each other.

Inerestingly, algebra (43) is a special case of the deformed algebra
$$d_\mu^{(n)}d_\nu^{+(n)}-\varphi(N_n)d_\nu^{+(n)}d_\mu^{(n)}
=-\eta_{\mu\nu}g(N_n),$$
$$[d_\mu^{(n)},d_\nu^{(n)}]=[d_\mu^{+(n)},d_\nu^{+(n)}]=0,$$
$$[d_\mu^{(n)},d_\nu^{+(m)}]=0, \quad n \neq m, \eqno(46)$$
the latter being a generalization of the deformed algebra suggested
recently [15] for the one-dimensional case.

Recently a very interesting possibility for the solution of the problem of
spectrum in dual models has been suggested [16]. These authors use the general
definition of duality
(9) without applying to the technique of $q$-deformations.

In conclusion we note that all the attempts to deform the Veneziano model
(harmonic oscillator) imply tree diagrams, or Born amplitudes, to be
unitarized (see the beginning of this paper). In an alternative approach
to the dual theory called dual amplitudes with Mandelstam analyticity [17],
the input scattering amplitude already has analytic properties expected from
planar unitarization (i.e. broad resonances, Mandelstam analyticity etc.).

To find the relevant operator formalism (deformed algebra?) is a challenge
for the theory.

We thank A.Bugrij and B.Struminsky for useful discussions.

The work of A.M. was supported in part by a Soros Foundation grant awarded
by the American Physical Society and by International Science Foundation grant.

\bigskip

\begin{center}
\bf{REFERENCES}
\end{center}
\bigskip
\begin{enumerate}
\item G.Veneziano, Nuovo Cimento, {\bf 57A} (1968) 190.
\item S.Fubini, D.Gordon and G.Veneziano, Phys. Letters, {\bf 29B} (1969) 679.
\item D.D.Coon and M.Baker, Phys. Rev., {\bf D2} (1970) 2349.
\item D.Coon, S.Yu and M.Baker. Phys.Rev. {\bf D5} (1972) 1429.
\item E.Gremmer and J.Nuyts, Nuc. Phys., {\bf B26} (1971) 151.\\
      J.M.Golden, Lett. Nuovo Cim., (Ser.2) {\bf 1} (1971) 893.\\
      S.Machida, Progr. Theor. Phys., {\bf 47} (1972) 2015.\\
      L.J.Romans, in Proceedings of the High Energy Physics Cosmology
      Conference, Trieste, (1989), eds. J.C.Pati et al.,
      World Scientific (1990).
\item M.Chiachian, J.F.Gomes and P.Kulish, Phys. Letters, {\bf B311} (1993) 93.
\item L.L.Jenkovszky, A.V.Mishchenko, B.V.Struminsky. Preprint ITP-93-36E and
      in Proc. of the HADRONS-93.
\item O.W.Greenberg, Phys. Rev. Lett., {\bf 64} (1990) 705;
      Phys. Rev., {\bf D43} (1991) 4111.\\
      R.N.Mohapatra, Phys. Lett., {\bf B242} (1990) 407.
\item S.V.Shabanov, Phys. Lett., {\bf B293} (1992) 117;
      J. Phys., {\bf A26} (1993) 2583.
\item M.B.Green, J.H.Schwarz and E.Witten, Superstring Theory (Cambridge Univ.
      Press, Cambridge, 1987).
\item E.T.Whittaker, G.N.Watson, A Course of Modern Analysis (Cambridge, At the
      University Press, 1927).
\item M.Kibler, in Symmetry and Structural Properties of Condensed Matter,
      ed. W.Florek, D.Lipinski and T.Lulek,
      World Scientific: Sigapore (1993) 445;
      Miscellaneous Physical Applications of Quantum Algebras,
      Preprint LYCEN/9358.
\item D.Ellinas, J. Phys., {\bf A26} (1993) L543;
      On Coherent States and q-Deformed Algebras,
      Preprint FTUV/93-37 and IFIC/93-20.
\item M.Chiachian, J.F.Gomes and R.Gonzalez Felipe, New Phenomenon of Nonlinear
      Regge Trajectory and Quantum Dual String Theory,
      Preprint HU-SEFT R 1994-05.
\item S.Meljanac, M.Milekovi\'c and S.Pallua, Unified View of Deformed
      Single-Mode Oscillator Algebras, Preprint RBI-TH-3/94 and PMF-ZTF-3/94
      (to appear in Phys. Lett {\bf B}).
\item D.B.Fairlie, J.Nuyts, A Fresh Look at Generalized Veneziano
      Amplitudes, Preprint DTP/94/19.
\item A.I.Bugrij, G.Cohen-Tannoudji, L.L Jenkovszky, N.A.Kobylinsky,
      Fortschritte Phys., {\bf 21} (1973) 427.
\end{enumerate}
\end{document}